\documentclass{vgtc}                          

\graphicspath{{figures/}{Figures/}{pictures/}{images/}{./}} 

\usepackage{times}                     

\usepackage{tabu}                      
\usepackage{booktabs}                  
\usepackage{lipsum}                    
\usepackage{mwe}                       
\usepackage{tabularx}
\usepackage{listings}
\usepackage{multirow}

\usepackage{mathptmx}                  

\usepackage{cite}
\usepackage{amsmath,amssymb,amsfonts}
\usepackage{algorithmic}
\usepackage{graphicx}
\usepackage{float}
\usepackage{textcomp}
\usepackage{xcolor}
\usepackage{xspace}
\usepackage[caption=false]{subfig}
\usepackage{enumitem}

\newcommand{\sysname}{PRISM-XR\xspace}

\onlineid{1920}

\vgtccategory{IEEE VR 2026 Papers}

\vgtcinsertpkg

\title{PRISM-XR: Empowering Privacy-Aware XR Collaboration\\with Multimodal Large Language Models}

\author{Jiangong Chen\thanks{co-first author. e-mail: jiangong@psu.edu} %
\and Mingyu Zhu\thanks{co-first author. e-mail: mintrrey@psu.edu} %
\and Bin Li\thanks{e-mail: binli@psu.edu}}
\affiliation{\scriptsize Department of Electrical Engineering, The Pennsylvania State University, University Park, PA 16802, USA}

\teaser{  
  \centering
  \includegraphics[width=\linewidth]{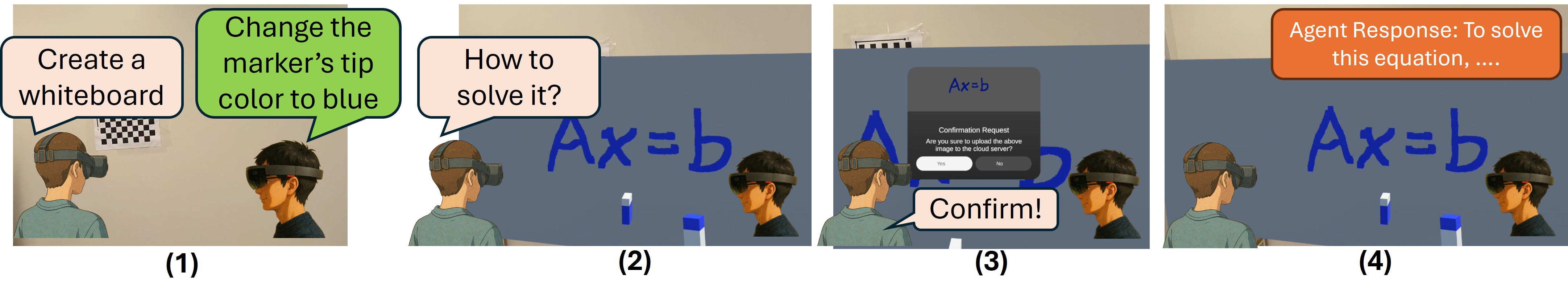}
  \caption{An example of multi-user collaboration in \sysname. In this example, two users collaborate with each other in the same physical space to create a virtual whiteboard, edit object properties, and leverage the Multimodal Large Language Model (MLLM) agent to help resolve a math problem. The users simply speak their requests, and \sysname will automatically execute commands and synchronize the virtual objects. To preserve privacy, it automatically crops the captured frame and requests user confirmation before uploading the cropped frame to cloud-based MLLM.
  }  
  \label{fig:teaser}
}

\abstract{
    Multimodal Large Language Models (MLLMs) enhance collaboration in Extended Reality (XR) environments by enabling flexible object and animation creation through the combination of natural language and visual inputs. However, visual data captured by XR headsets includes real-world backgrounds that may contain irrelevant or sensitive user information, such as credit cards left on the table or facial identities of other users. Uploading those frames to cloud-based MLLMs poses serious privacy risks, particularly when such data is processed without explicit user consent. Additionally, existing colocation and synchronization mechanisms in commercial XR APIs rely on time-consuming, privacy-invasive environment scanning and struggle to adapt to the highly dynamic nature of MLLM-integrated XR environments. In this paper, we propose \sysname, a novel framework that facilitates multi-user collaboration in XR by providing privacy-aware MLLM integration. \sysname employs intelligent frame preprocessing on the edge server to filter sensitive data and remove irrelevant context before communicating with cloud generative AI models. Additionally, we introduce a lightweight registration process and a fully customizable content-sharing mechanism to enable efficient, accurate, and privacy-preserving content synchronization among users. Our numerical evaluation results indicate that the proposed platform achieves nearly 90\% accuracy in fulfilling user requests and less than 0.27 seconds registration time while maintaining spatial inconsistencies of less than 3.5 cm. Furthermore, we conducted an IRB-approved user study with 28 participants, demonstrating that our system could automatically filter highly sensitive objects in over 90\% of scenarios while maintaining strong overall usability.
} 


\begin{document}

\firstsection{Introduction}

\maketitle
Integrating Multimodal Large Language Models (MLLMs) into Extended Reality (XR) applications (see \cite{srinidhi2024xair,xiu2024lobstar,xiu2025viddar,zhu2026genai}) revolutionizes user collaboration in immersive environments. In recent years, social XR applications such as VRChat and Meta Horizon Worlds have gained significant attention, driven by the rising popularity of the metaverse. The integration of MLLMs enhances these experiences by enabling users to collaboratively build and modify virtual worlds in real-time (see \cite{davies2024roblox,earle2024dreamcraft,merino2023interactive}), enriching interaction and fostering stronger social connections. Beyond virtual collaboration, MLLMs also enhance contextual understanding of the physical environment (see \cite{srinidhi2025xr,chen2025llmer}), allowing for seamless interaction when users share the same physical space. For instance, in an office workspace or home environment, multiple users wearing XR headsets can dynamically customize surroundings using voice commands. One person might say, ``\emph{Add a virtual whiteboard here and some colored pens}." Another user could start drawing on the whiteboard and then prompt, ``\emph{Summarize our writing and generate digital objects based on the drawing}." Instantly, the system responds by placing interactive objects in the shared space, enabling all participants to engage with and modify the digital overlays in real-time.

Despite its potential, integrating MLLM into collaborative XR applications faces several challenges: i) The visual inputs captured by XR devices may contain sensitive user information (see \cite{yamakami2020privacy,abraham2022implications}),
such as credit cards left on the table or facial identities of other users, especially in physical environments where users interact face-to-face or in private spaces. Uploading such data to cloud servers managed by commercial entities raises significant privacy concerns (see \cite{hu2021lenscap,corbett2023bystandar}), particularly without explicit user consent. ii) Real-world environments often contain numerous irrelevant objects, making it difficult for MLLMs to accurately extract the necessary context (see \cite{shi2023large,huang2025breaking}) to fulfill user requests. iii) Seamless interaction with shared digital objects in a physical space requires precise coordinate alignment among users (see \cite{you2020range,mecheri2016evaluation}).
However, existing colocation solutions either require time-consuming and privacy-invasive environment scanning (see \cite{metassa,sanchorsharing,gcloudanchor}) or have prohibitive constraints of keeping fiducial markers visible (see \cite{olson2011apriltag,garrido2014automatic}). 
iv) The highly dynamic nature of MLLM-driven XR applications poses compatibility challenges (see \cite{kobenova2024social}) for traditional network synchronization mechanisms, which are primarily designed for predefined scenes and user actions.

As such, the following research questions arise: i) How can visual inputs be pruned to protect user privacy and reduce contextual noise while still providing sufficient information for seamless interaction? ii) Is there a low-cost, efficient, accurate, and privacy-aware colocation strategy that ensures an immersive collaboration experience for physical users? iii) How can objects and animations in XR environments be efficiently synchronized across users while considering the highly dynamic properties of MLLM-powered XR applications?
Prior research (see \cite{chen2024gpt, de2024llmr, giunchi2024dreamcodevr}) has explored various approaches to integrating MLLMs into XR applications, primarily through script generation or structured data. 
However, these studies have not addressed the risk of privacy leakage when communicating with cloud-based MLLM servers.
Although some studies (see \cite{qian2022boosting, corbett2023bystandar, hu2023magiccloth}) have proposed methods to obscure or replace sensitive objects to protect user privacy in XR environments, these approaches may lead to missing or altered environmental contexts, which can hinder effective MLLM integration.

In this paper, we propose \sysname, a \underline{PRI}vacy-aware and \underline{S}hared platform that integrates \underline{M}LLMs into \underline{XR} environments.
\sysname enables multiple users to collaborate seamlessly by creating virtual objects and animations that interact with real-world environments through simple voice commands. By leveraging local context collection, privacy-aware processing on edge servers, and powerful closed-source commercial cloud Artificial Intelligence (AI) models, \sysname processes user voice inputs in a multi-stage pipeline, fusing multimodal data to generate structured JSON outputs that translate into XR tasks. To protect privacy and remove contextual noise, \sysname either converts sensitive visual inputs into text-based descriptions or explicitly seeks user confirmation before uploading cropped frames to the cloud. For users in the same physical space, a one-time, sub-second registration process ensures seamless synchronization of all user actions in the rest of the playing session.
\sysname serves as a simplified XR interaction and collaboration platform designed for educated lay users with no prior experience in programming or XR technologies.

The key design aspects of \sysname are summarized as follows. First, \sysname employs a state-of-the-art object detection model on the edge server to analyze the user's Field of View (FoV) and generate a textual summary, which replaces raw image frames in the prompts sent to the MLLM. If the generated description is insufficient to fulfill user requests, \sysname will specify a cropped area based on the description and request user confirmation before uploading the cropped section. Second, \sysname fuses local XR tracking systems of commercial XR devices with a marker-based tracking approach through a lightweight registration process. Users can freely explore the physical environment without relying on fiducial markers, which are used only during initialization to facilitate virtual content alignment. Lastly, \sysname features a lightweight and fully customizable synchronization mechanism tailored to MLLM responses and XR user interactions. XR-related objects and properties are stored on the edge server and synchronized locally to reduce latency and bandwidth consumption.

The main contributions of our paper are:
\begin{itemize}
    \item We propose a privacy-aware frame processing method that preserves user privacy by automatically excluding sensitive information from being uploaded to the cloud server.
    \item We introduce lightweight and customizable techniques for spatial alignment of users and virtual content synchronization, which enables effective multi-user collaboration in MLLM-powered XR applications.
    \item We develop and open-source\footnote{\url{https://github.com/SNeC-Lab-PSU/PRISM-XR}} a fully deployable end-to-end system that chains all components. The system features a user-friendly interface with natural language input and is designed to be extensible and ready for real-world deployment.
    \item We conduct both system-level numerical evaluations and a preliminary user study to assess the usability of \sysname, which reveals its powerful functionality and indicates opportunities for further optimization.
\end{itemize}

The remainder of this paper is organized as follows: \Cref{sec:ref} summarizes existing literature in related fields. \Cref{sec:overview} presents the overview and workflow of our proposed system. \Cref{sec:design} showcases a detailed description of the key design aspects of our system. \Cref{sec:implementation} provides the software and hardware implementation of our system. \Cref{sec:num:study} and \Cref{sec:user:study} describe the evaluations on both system performance and usability with human subjects, and \Cref{sec:conclusion} concludes the paper.

\section{Related Work}\label{sec:ref}
In this section, we summarize the relevant literature in four key areas: LLM Agents, LLMs in XR, multi-user XR applications, and user privacy protection in XR.

\textbf{LLM agents.}
With recent advances in MLLMs, AI agents have become a popular research topic. Researchers have explored various approaches to leveraging MLLMs for automatically and intelligently handling complex tasks without human intervention. Some prior studies (see \cite{park2023generative, wang2023voyager}) have experimented with LLMs in video games, where Non-Player Characters (NPCs) exhibit believable human-like behavior. Other works (see \cite{brohan2023rt, wang2023prompt, yu2023language}) incorporate AI agents into end-to-end robotic control, enabling emergent semantic reasoning or interactive robotic behaviors. Meanwhile, more recent research (see \cite{schmidgall2025agent,boiko2023emergent, li2024mlr}) develops AI agents to accelerate scientific discovery, such as conducting autonomous literature reviews, proposing research ideas, and planning experiments. 

\textbf{LLMs in XR.}
Recent studies have also shown emerging trends in applying LLM agents to XR environments, which immerse users by enabling interaction with virtual objects that go beyond physical constraints.
Some pioneer works (see \cite{de2024llmr,chen2025llmer,chen2024gpt,giunchi2024dreamcodevr,zhu2026genai}) have introduced AI agents into XR environments; however, these primarily rely on textual context and generate code scripts or JSON data to control interactive objects or animations. Other works (see \cite{srinidhi2024xair,xiu2024lobstar}) explore real-world image inputs by leveraging MLLMs, but they do not address user privacy issues, as user-captured images are shared directly without notification. Moreover, most of these studies target single-user interaction and overlook the challenges and opportunities of multi-user XR applications, such as colocation issues and efficient synchronization. A more recent study (see \cite{kobenova2024social}) investigates multi-user collaboration in designing VR worlds with LLMs, but it does not consider AR/MR scenarios, particularly those where users are located in the same real-world space.
For a broader review of XR and AI integration, readers can refer to \cite{hirzle2023xr,tang2025llm}.

\textbf{Multi-user XR applications.}
Beyond the interaction between the user, AI agent, and XR environment, XR-based collaboration introduces an additional layer of immersion in a wide range of application scenarios. 
Researchers have demonstrated the effectiveness of multi-user XR systems in areas like collaborative construction \cite{yao2022scalable}, firefighter training \cite{dong2023collaborative}, and remote learning (see \cite{chen2021motion, chen2022enhancing}). However, these studies do not explore the potential benefits of integrating MLLM agents into collaborative XR applications. In addition, multi-user collaboration within real-world environments poses greater localization challenges than single-user scenarios. Each user may have different perspectives and movements, making it difficult to merge individual maps and maintain accurate localization in a shared environment. Some works (see \cite{li2017corb, liu2021edgesharing, dhakal2022slam}) address these issues by focusing on efficient global map construction and reducing communication latency between multiple users and the server. Others (see \cite{li2020promar, ran2020multi}) investigate methods for exchanging visual features or keyframes to align users’ coordinate systems. Nevertheless, these studies concentrate on camera localization and do not delve into the dynamics of MLLM-driven XR collaboration, particularly the synchronization of virtual objects or animations.

\textbf{User privacy protection in XR.}
With the integration of AI-driven analysis and inter-personal interaction, privacy concerns in XR have attracted researchers' attention\cite{rajaram2023eliciting}.
Prior studies highlight risks such as bystander data leaks and hidden machine-learning operations that extract personal information without user awareness \cite{o2023privacy, lehman2022hidden}, while concerns about unwanted exposure and perception bias of reality further hinder user adoption of XR system\cite{lammerding2021too}. Recent privacy-enhancing solutions, including eye gaze tracking and spatial awareness for selective bystander protection \cite{corbett2023bystandar} and object replacement through patterned tracking for hiding sensitive items in XR live streaming \cite{hu2023magiccloth}, mitigate these risks through on-device processing, selective masking, and secure inference pipelines.
However, integrating MLLMs in collaborative XR introduces new privacy challenges \cite{yao2024survey}, as these models typically rely on cloud-based processing, potentially uploading sensitive visual data without explicit user control. This shift from local processing
to remote server necessitates new strategies for safeguarding privacy in MLLM-powered XR environments.
For a more comprehensive review of privacy concerns in XR environments, we recommend readers refer to \cite{rajaram2023eliciting, alkaeed2024privacy, hadan2024privacy}.

\section{System Overview}\label{sec:overview}
\begin{figure}[hbtp]
    \centering
    \vspace{-0.1in}
    \includegraphics[width=0.45\textwidth]{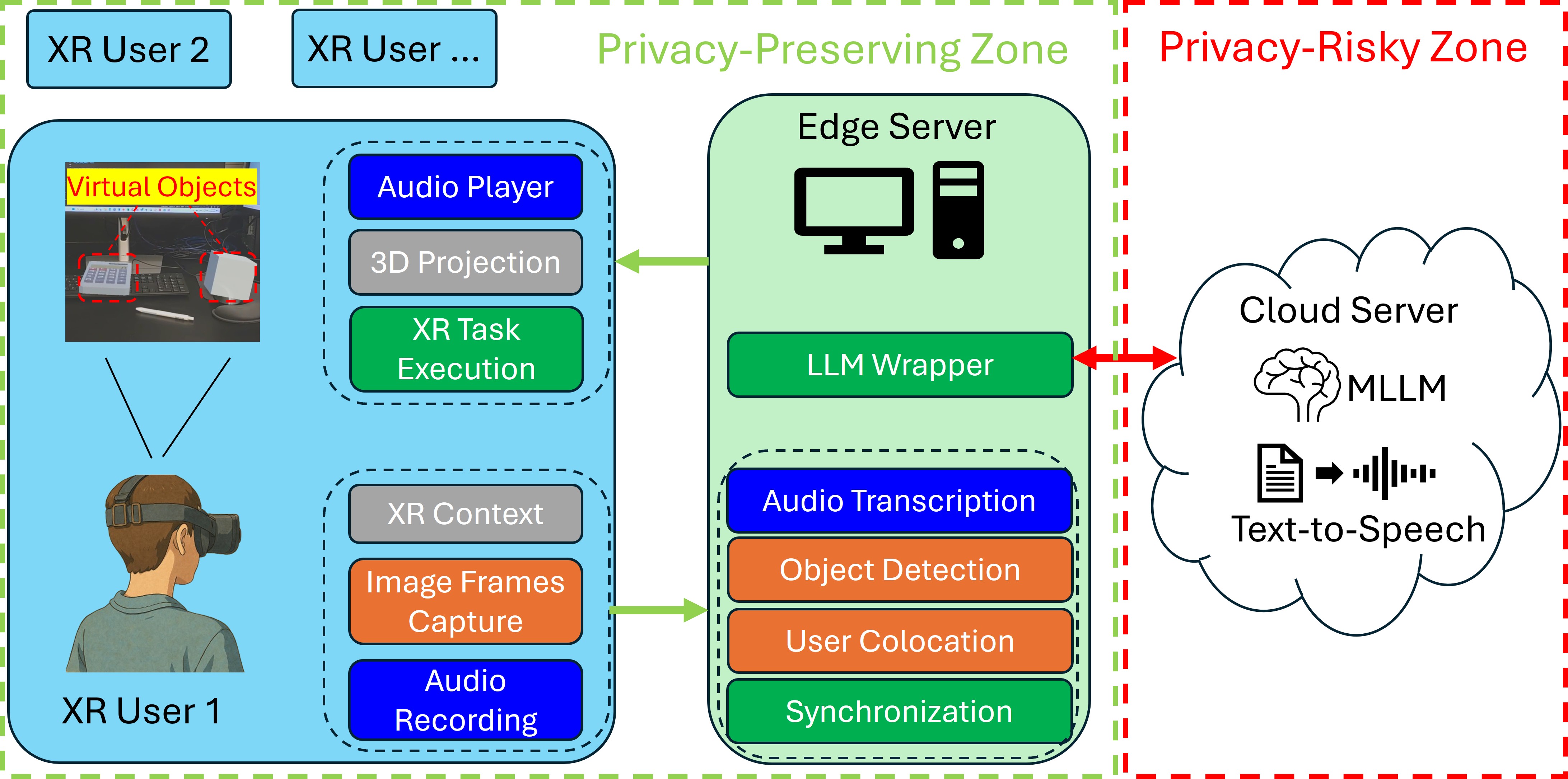}
    \vspace{-0.1in}
     \caption{System architecture of \sysname.}     
     \vspace{-0.1in}
    \label{fig:sys-arch}
\end{figure}

In this section, we provide an overview of \sysname, a framework that enables multiple users to collaborate within the same physical space to create virtual objects and animations using voice inputs. \sysname extends the LLMER\cite{chen2025llmer} framework to support multi-user interactions and introduces novel methods for incorporating real-world image inputs to enhance scene understanding. The system architecture of \sysname, including data collection on the client side and multi-layered processing on edge and cloud servers, is illustrated in \cref{fig:sys-arch}.

Unlike existing systems that rely on cumbersome manual inputs \cite{de2024llmr}, gesture-based audio triggers \cite{chen2025llmer}, or controller-based audio activation \cite{giunchi2024dreamcodevr}, \sysname employs a more flexible keyword-activated approach.
Users initiate their requests by starting with a predefined keyword, which triggers automatic audio recording when the keyword is detected and stops when silence is identified. The recorded audio segments are transcribed using the OpenAI Whisper model \cite{radford2023robust}. To ensure user privacy, all processes are handled locally: keyword detection and audio recording are performed directly on the client device, while audio transcription is executed on the edge server. This design eliminates the need to send sensitive audio data to external services.
Additionally, the XR device captures a frame reflecting the user’s field of view (FoV) along with the camera pose at the time of capture. Instead of uploading raw frames to the cloud, the edge server employs a privacy-conscious approach
by either extracting a textual description of the frame to incorporate into the prompt or cropping the frame and requesting user consent before incorporating the cropped frame into the prompt.
Once the user's voice commands and contextual information are collected, the edge server communicates with the cloud AI server to process the request and parse the JSON outputs from the cloud for reliable rendering on client devices.

To facilitate multi-user collaboration, \sysname requires users to undergo a lightweight registration process. This process uses a semi-marker-based method to efficiently and accurately align users’ coordinate systems while maintaining flexibility for users to navigate real-world environments without FoV constraints. All responses from MLLMs are broadcast across registered users, allowing them to observe objects and animations rendered in the same physical space simultaneously. \sysname also ensures seamless synchronization of interactable objects, such as grabbable cubes, by employing ownership-aware and change-driven data communication. This design supports collaborative actions under the immersive XR environments, such as the handover of virtual objects. Considering the possible keyword conflicts when users are within the same physical space, different keywords are assigned to each user. In the next section, we will elaborate on the detailed design principles for our system.

\section{System Design}\label{sec:design}
In this section, we discuss three key design principles of \sysname that empower multi-user collaboration using MLLMs with real-world visual inputs. First, we describe the privacy-aware frame processing of \sysname, which is essential for providing concise and secure real-world context to MLLMs. Next, we outline the semi-marker-based user registration process, designed to ensure efficient, accurate, and flexible user colocation. Finally, we present \sysname's content synchronization mechanism, tailored to meet the demands of dynamic environments powered by MLLMs.

\subsection{Privacy-aware Frame Processing}\label{sec:frame-process}
\begin{figure}[hbtp]
    \centering
    \vspace{-0.1in}
    \includegraphics[width=0.95\linewidth]{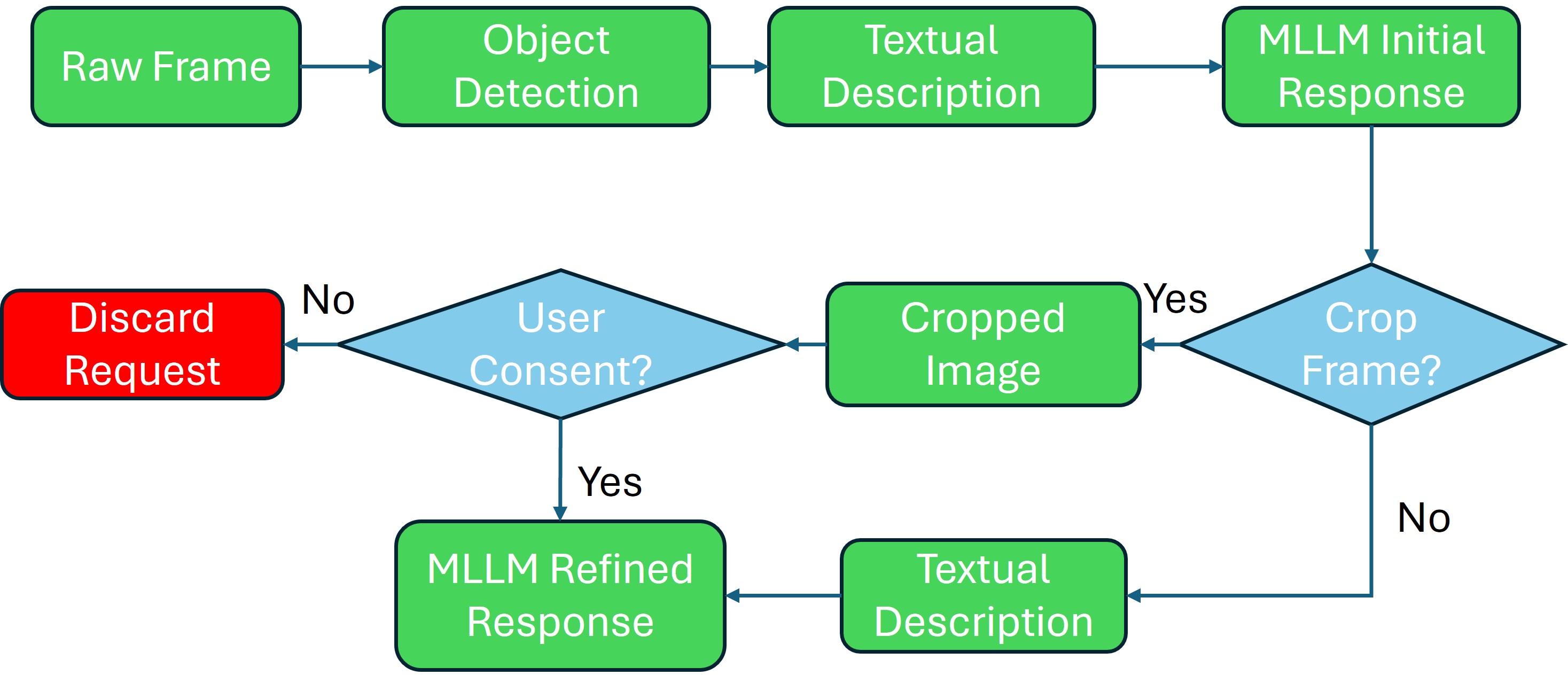}
    \caption{Workflow of privacy-aware frame processing.}
    \vspace{-0.1in}
    \label{fig:privacy-workflow}
\end{figure}

To enable more fine-grained interactions in XR worlds, visual inputs for MLLMs are essential, as they provide contextual information beyond coarse textual descriptions. However, real-world environments often include sensitive information, particularly in multi-user collaboration scenarios. For instance, when one user faces another during an interaction with a virtual chessboard, the captured screenshot may include a person’s face, which raises privacy concerns if uploaded to cloud servers \cite{corbett2023bystandar, hu2021lenscap}. To address the privacy concern, \sysname proposes a privacy-aware frame processing mechanism with automatic frame cropping and explicit user consent requests, with the workflow depicted in \Cref{fig:privacy-workflow}.

\begin{figure*}[hbtp]
    \centering
    \begin{minipage}[c]{0.74\textwidth}
        \centering
        \subfloat[YOLO detection results of raw image containing sensitive objects.]{\includegraphics[width=0.3\linewidth]{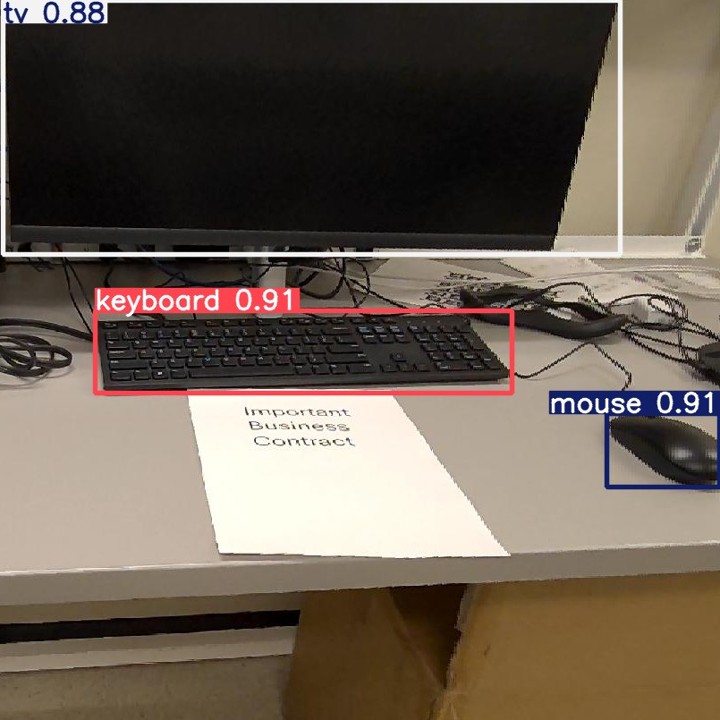}\label{fig:yolo}}
        \hfill
        \subfloat[Collider of real-world environment and raycasting.]{\includegraphics[width=0.3\linewidth]{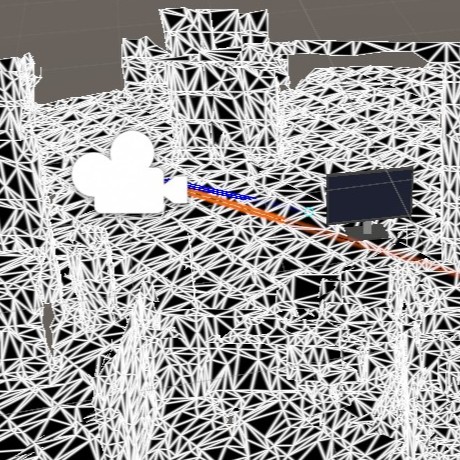}\label{fig:raycast}}
        \hfill
        \subfloat[Confirmation dialog, where sensitive objects are excluded.]{\includegraphics[width=0.3\linewidth]{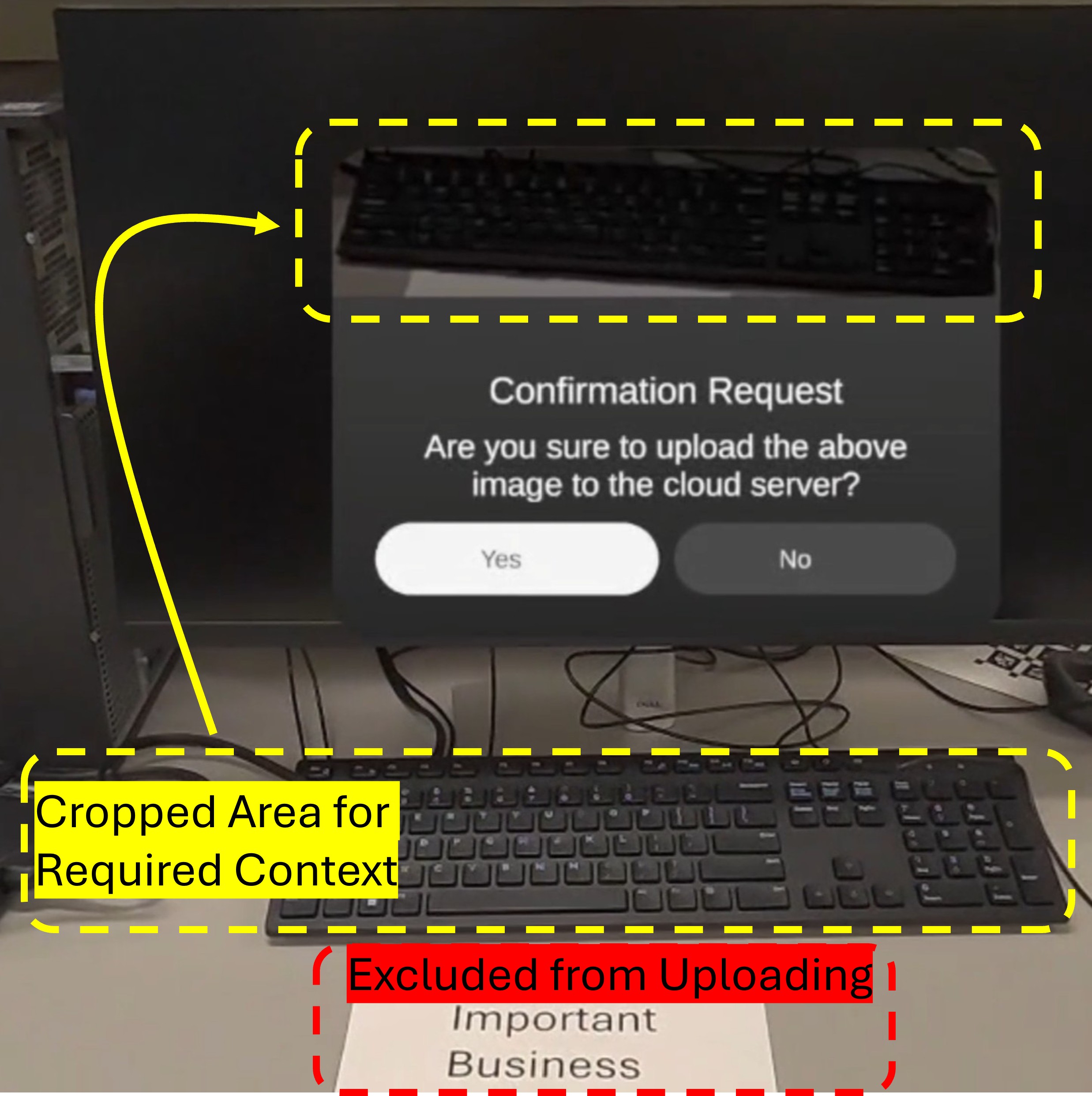}\label{fig:confirm}}
        \vspace{-0.1in}
        \caption{Examples of critical steps in privacy-aware frame processing.}
        \label{fig:ex-workflow}
    \end{minipage}
    \hspace{0.1in}
    \begin{minipage}[c]{0.222\textwidth}
        \centering
        \includegraphics[width=\textwidth]{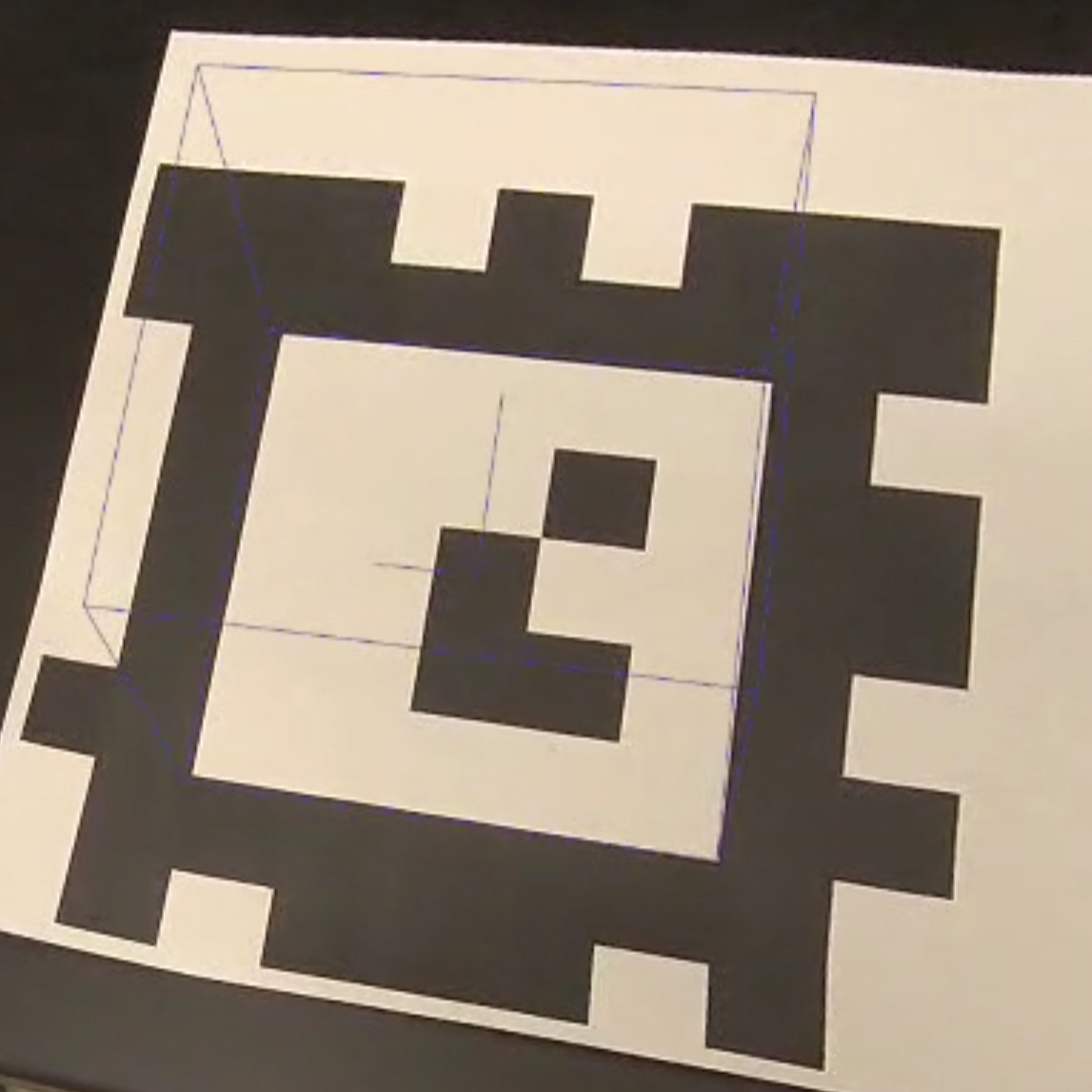}
        \caption{Successful user registration with a wire-frame cube.
        }
        \label{fig:registration}
    \end{minipage}
    \vspace{-0.2in}
\end{figure*}

Instead of uploading raw frames captured by XR devices to the cloud server, \sysname converts images into textual descriptions on the edge server to safeguard user privacy.
The system employs the state-of-the-art YOLO v11 model \cite{Jocher_Ultralytics_YOLO_2023}, chosen for its high accuracy, generalizability, and broad adoption across scenarios. The model’s output includes the detected object’s name, center pixel coordinates, bounding boxes, and confidence levels. An example detection result is illustrated in \Cref{fig:yolo}, and the textual description of the user's FoV follows this format:
 
\begin{lstlisting}
keyboard, center (327.80, 352.12), box (66.47, 66.03, 589.13, 638.21), confidence 0.91
\end{lstlisting}
This data is incorporated into the prompt for MLLMs to provide relevant real-world contexts.
For example, the MLLM is aware of the user's body gestures when the hand is detected, enabling vague requests like ``moving objects here".
Since object detection outputs only pixel coordinates within the image frame, additional processing is required to calculate accurate 3D world positions.
Similar to \cite{srinidhi2024xair}, \sysname records the camera pose at the time of frame capture and casts a ray from the camera position along the direction derived from the pixel coordinates. The ray will hit the colliders for the real-world environment, which is constructed dynamically by XR devices such as Microsoft HoloLens and Meta Quest 3. The hit point is then set as the initial position of the target object. To enhance the natural appearance of rendered objects, \sysname applies pose correction. This process adjusts the object's position based on its size to ensure proper surface attachment and aligns its orientation with the surface's normal. An example of creating objects from pixel coordinates on a HoloLens device is provided in \Cref{fig:raycast}. Note that creating objects or animations in the world coordinates, as proposed in LLMER \cite{chen2025llmer}, is also supported in \sysname without involving this frame processing.

In scenarios where textual descriptions alone are insufficient, \sysname leverages MLLMs to automatically crop the frame. This is achieved by incorporating the textual description into the prompt and requesting the model to return a ``CropArea" property that specifies the region to be cropped. MLLM will return an array indicating the area if it is necessary or simply ``None" if the request does not need frame information.
\Cref{fig:confirm} illustrates an example in which the user requests color information about a keyboard, which cannot be captured by the textual description alone. In this case, the YOLO model detects the keyboard’s position, the MLLM identifies the relevant cropping area, and \sysname generates a cropped frame that excludes sensitive content for user approval.
Only after the user verifies that the cropped image contains no sensitive information is it included in the prompt sent to the cloud server, while the un-cropped frame is only processed by the edge server and will not be uploaded to cloud.
Note that there are some alternative approaches to process the frame, such as blurring or grey-scaling the sensitive objects. However, those approaches require an accurate detection of sensitive objects, which typically requires fine-tuning for a more personalized detection. In comparison, our approach leverages the power of MLLMs to intelligently determine the crop area with the help of YOLO models detecting common-sense objects without finetuning.
Beyond protecting user privacy, the cropped frame also eliminates irrelevant contextual information, keeping the MLLM focused on essential details. For example, the cropped image excludes extraneous elements such as tables and chairs, reducing distractions for the MLLM. Additionally, commercial APIs often calculate token consumption for image inputs based on pixel count, meaning that cropped images can significantly reduce token costs.

\subsection{Semi-Marker-Based User Registration}\label{sec:registration}
A key requirement for multi-user XR applications, particularly those involving real-world backgrounds and multiple users in the same physical location, is the alignment of users' coordinate systems. This alignment ensures that virtual objects are perceived as being attached to the same physical spaces, regardless of individual users' perspectives. However, commercial APIs like Meta Shared Spatial Anchors (SSA)\cite{metassa}, Microsoft Spatial Anchor Sharing \cite{sanchorsharing}, and Google Cloud Anchors \cite{gcloudanchor} require users to scan environments and upload data to cloud servers. This process is not only time-consuming but also raises significant privacy concerns.

In contrast, \sysname employs a lightweight and privacy-aware method for aligning users' coordinate systems.
This approach integrates the tag-based localization system, AprilTag \cite{olson2011apriltag}, to facilitate an efficient registration process for each user. 
Although AprilTag is a well-established technology, our main contribution lies in its seamless integration with commercial devices' internal localization algorithms. Specifically, we use AprilTag solely for aligning users' world coordinate systems without overriding HMDs' original localization capability. Particularly, users identify nearby tags in fixed locations and say ``registration'', by which \sysname captures a screenshot and applies the AprilTag detector to identify tags and estimate their poses in the user's world coordinate system.
If the tag is successfully identified, the edge server stores the tag pose as the user’s registration information, sends the pose back to the user, and renders a wireframe cube at the tag position on the user's XR device, as shown in \Cref{fig:registration}. If the tag is not detected, the edge server notifies the user of the registration failure and suggests attempting the process again.
Once registration is completed, an object's pose in a user's coordinate system can then be transformed to the other user's coordinate system by straightforward basis change and calculation, while the user's movements will not be limited to the AprilTag's constraints, like keeping the tags within sight.
Note that the registration is required each time the application starts, and all processing is handled locally without a cloud server to preserve user privacy.
Additionally, if a user observes that synchronized objects from other users appear misaligned, they can repeat the registration process at any time to correct discrepancies.

\subsection{Content Synchronization}\label{sec:sync}
Beyond coordinate transformation, a multi-user collaboration system must also ensure the timely synchronization of virtual content in the scene, such as the creation or modification of objects. Instead of relying on commercial APIs like Photon Fusion — which upload and store user data in cloud servers — we leverage the edge server and design a customized synchronization protocol tailored to the dynamic environments powered by MLLMs. This approach ensures efficient content sharing while maintaining privacy and minimizing latency. We categorize content synchronization into two main types: MLLM responses and user interactions, each requiring distinct synchronization strategies.

\subsubsection{Synchronizing MLLM Responses}
Recall that all responses from the cloud AI server, such as those for object or animation creation, are structured as JSON data. These responses can be broadcast to all users to ensure consistency in the virtual world. Since JSON strings are relatively short, they incur minimal bandwidth usage and latency. However, some MLLM responses, such as creating an object at a specific position within a user’s FoV, depend on user-specific contexts that are not available for other users.
To address this, we define three coordinate spaces for the poses of MLLM-generated virtual content:
i) World Space: refers to the absolute 3D pose in the world coordinate system of the user who made the request. ii) Local Space: for objects attached to other objects as children or animations operated within the local coordinate system. iii) Pixel Space: works in conjunction with the raycasting process, as discussed in \Cref{sec:frame-process} and illustrated in \Cref{fig:raycast}.
The user who creates the virtual content is designated as the owner user.
Each MLLM response will be sent to the owner user, who converts the content's pose from local or pixel space to world space, and then broadcasts to other users, applying the necessary coordinate transformations.

\subsubsection{Synchronizing User Interactions}
User-driven modifications, like pose updates of interactable objects created by users, require a separate synchronization mechanism.
\sysname employs a change detection strategy to actively monitor and synchronize changes to virtual object properties. Updates are only sent when changes exceed a predefined threshold, and the time since the last synchronization surpasses a specified interval. Those parameters are set to balance the bandwidth consumption and smoothness of synchronization.
In multi-user scenarios, conflicts may arise when multiple users interact with the same virtual object. To resolve this, \sysname distinguishes between owned objects and synced objects:
i) Owned objects: objects a user directly interacts with. Their properties are uploaded by the client to the edge server. ii) Synced objects: objects not directly owned by the user. Their properties are updated by receiving data from the edge server.
When a user interacts with an object, it is automatically added to their list of owned objects and removed from the synced objects list. Simultaneously, the edge server notifies the previous owner to update their lists accordingly.

To enhance communication efficiency, \sysname uses a specialized data structure for synchronizing interactable objects instead of relying on JSON strings. This structure encodes properties into raw bytes, reducing the size of synchronization messages to only 48 bytes per object among registered users (4 bytes for object id, 12 bytes for 3D positions, 16 bytes for 3D rotations in quaternion, 12 bytes for 3D scales, and 4 bytes for special events handling).
Besides, we use native socket and local coordinate transformation to reduce the communication overhead further.

\section{System Implementation}\label{sec:implementation}
In this section, we detail the software and hardware implementation of \sysname on the client side, the edge server, and the physical environment used for evaluations.

\textbf{Client.}
The client application for \sysname is developed as a Unity project using C\#. It is built upon the open-sourced framework LLMER \cite{chen2025llmer}. Note that the original LLMER framework only supports the Meta Quest series, while our platform supports both Meta Quest devices and Microsoft HoloLens 2. In the HoloLens version, we integrate Microsoft Mixed Reality Toolkit 3 (MRTK3) to enable XR interactions, such as grabbing virtual objects. Screenshots are captured using the PhotoCapture API from the Unity.Windows package.
This API provides the camera pose and projection matrix for the captured frames.
The keyword activation mechanism is implemented through MRTK3's KeywordRecognitionSubsystem, while audio recording and silence detection are managed by Unity’s AudioModule. The initial audio recording duration is set to 2 seconds and extends in 1-second increments until the audio magnitude consistently falls below a predefined threshold. All these processes are executed offline directly on the client device.
In the Quest version, we use Meta XR All-in-One SDK to build the basic XR interactions. The Passthrough Camera API within the SDK is used to capture the user's FoV and facilitate raycasting in the real-world environment. We use the Porcupine Unity package to implement the keyword activation while using the same recording and silence detection mechanism as on HoloLens.
On both devices, socket communication with the edge server is established using the System.Net.WebSockets package, by which JSON strings and serialized bytes mentioned in \Cref{sec:sync} are transmitted through bi-directional TCP connections.
To ensure smooth interactions among users and avoid bandwidth overload, the synchronization frequency of user interactions is set to 60 Hz.

\textbf{Edge server.} The edge server for \sysname is developed in Python and supports both Linux and Windows operating systems. It uses OpenAI's official Python API to interact with the cloud server to generate MLLM responses and the Text-to-Speech (TTS) model to generate synthetic voice. The structured outputs mode of the GPT-4o model ensures that responses conform to a specified JSON schema, enforcing constraints like requiring object positions to be arrays of three float numbers. However, the model’s accuracy in determining the exact position relies on the provided context and its inherent capabilities.
For privacy-aware audio transcription and frame processing, the server deploys offline OpenAI Whisper model \cite{radford2023robust} and YOLO v11 \cite{Jocher_Ultralytics_YOLO_2023} model. The socket server is built upon the WebSocket API in Python.

\textbf{Hardware and physical environment.} We have deployed the client application on both Microsoft HoloLens 2 running Windows 10 Holographic OS and Meta Quest 3 running Meta Horizon OS v74. Throughout our preliminary testing, we observed that user experience on Meta Quest devices comprehensively surpasses the experience using HoloLens. Unless otherwise specified, the experimental results are obtained from the client deployed on Meta Quest 3 devices, operating at a resolution of $2064\times2208$ pixels per eye.
The edge server runs on a Lambda workstation with the following specifications: Intel Core i9-10980XE CPU @3.00GHz × 36, NVIDIA GeForce RTX 3070 Graphics Card × 4, 128 GB memory, 4 TB disk, and Ubuntu 22.04 LTS. The edge server connects to a TP-Link Archer AX50 router via Ethernet, while client devices connect to the router through WiFi 5 (802.11ac). The server uses an additional network port for Internet access to communicate with the cloud generative AI server.
All experiments and user studies are conducted in a research lab (around 550 square feet) where the XR device, edge server, and router are co-located. Though most actions can be completed while still standing, users can move freely around the room for a more immersive experience.

\section{Numerical Study}\label{sec:num:study}
We first conduct a series of numerical studies to evaluate the system-level performance of \sysname, including the accuracy of generated responses from various MLLMs either in cloud or edge, colocation accuracy and efficiency, and synchronization latency.

\subsection{Accuracy of MLLM Responses}
\sysname involves two stages of communication with MLLMs: the \textit{initial stage} aims to parse and classify the request, obtain the crop area and corresponding categories; the \textit{refined stage} retrieves the structured response by integrating proper prompts based on the response in the initial stage.
We assess the accuracy of MLLM-generated responses for both stages and compare the performance of cloud-based models against open-source models running on the edge server. Specifically, we test OpenAI's o1, o3-mini, and GPT-4o as cloud-based models, and LLaVA 13B and Llama 3.2 Vision 11B deployed with the Ollama framework as edge-based models. Larger models exceed the capability of edge servers in general research labs that are not specifically dedicated to large-scale AI models, while smaller models struggle with structured output generation, often resulting in endless processing loops.
 
To evaluate the initial stage, we propose 15 typical user requests across different categories,
along with their associated real-world contexts (FoV descriptions as discussed in \Cref{sec:frame-process}). Seven of these requests require visual context from cropped images (e.g., translating text from a real-world book), while the remaining eight can be fulfilled using textual descriptions alone (e.g., moving objects towards detected objects in textual context). Among the cropped image requests, four explicitly specify the target area, while three are implicit. To ensure the generated requests align with practical user requests, we manually filter out the improper requests and refine the expected responses. Generating requests through LLMs with proper correction eliminates bias and enhances generality compared to directly proposing user requests. We leverage the following four metrics to evaluate the performance: Classification Accuracy, Crop Recall, Crop Fallout, and Generation Time.
Classification accuracy represents the proportion of correctly categorized requests. Crop Recall measures the proportion of cases where the Intersection over Union (IoU) between the returned and expected cropped areas exceeds 0.5, a sufficiently large value indicating that real-world visual contexts are properly provided. Crop Fallout captures instances where a cropped area is incorrectly generated when ``None" was expected, which may lead to unnecessary pop-ups. Ideally, a model should achieve high accuracy and Crop Recall while maintaining low Crop Fallout and Generation Time.

\begin{table}[hbtp]
\centering
\begin{tabular}{@{}c|ccc|cc@{}}
\toprule
\multirow{2}{*}{\textbf{Metric}} & \multicolumn{3}{c|}{\textbf{Cloud}} & \multicolumn{2}{c}{\textbf{Edge}} \\
 & \textbf{o1} & \textbf{o3-mini} & \textbf{GPT-4o} & \textbf{LLaVA} & \textbf{Llama V} \\
\midrule
\multicolumn{6}{l}{\textit{Initial stage to parse request, obtain crop area and categories}} \\
\midrule
CA (\%) & 88.67 & 72.67 & \textbf{89.33} & 30.67 & 23.33 \\
CR (\%)             & 71.43 & 72.86 & \textbf{78.57} & 1.43  & 7.14 \\
CF (\%)            & \textbf{0} & 5.00 & 1.25 & 95.00 & 15.00 \\
GT (s)          & 11.407 & 8.305 & \textbf{2.415} & 7.032 & 3.164 \\
\midrule
\multicolumn{6}{l}{\textit{Refined stage to retrieve structured response}} \\
\midrule
$r_{obj}$ (\%)            & 96 & 85 & \textbf{97} & 1 & 29 \\
$t_{obj}$ (s)              & 20.285 & 9.832 & \textbf{2.509} & 7.192 & 3.835 \\
$r_{ani}$ (\%)         & \textbf{100} & 100 & 99 & 0 & 1 \\
$t_{ani}$ (s)           & 9.708 & 9.929 & \textbf{3.611} & 10.345 & 4.469 \\
\bottomrule
\end{tabular}
\caption{Performance comparison across models. CA, CR, CF, GT are short for Classification Accuracy, Crop Recall, Crop Fallout, and Generation Time, respectively. $r_{obj}$ and $r_{ani}$ stand for the successful ratio of correct objection creation or animation creation response, while $t_{obj}$ and $t_{ani}$ stand for their consumed time, respectively.
}
\vspace{-0.2in}
\label{tab:eva:accuracy}
\end{table}

We use structured outputs for both OpenAI’s models and locally deployed models with Ollama to ensure compliance with a specified JSON schema. We use exactly the same prompts for all tested models and repeat each request 10 times to compute the average performance metrics. The results, as shown in \Cref{tab:eva:accuracy}, indicate that cloud models consistently outperform edge-based models in classification accuracy, Crop Recall, and Crop Fallout, emphasizing the necessity of cloud-based processing that raises significant privacy concerns to be addressed.
Interestingly, o1 and o3-mini demonstrate lower Crop Recall compared to GPT-4o, likely due to their tendency to return ``None" for cropped areas more frequently. Note that the effectiveness of Crop Recall is influenced by the level of detail in user requests—when users provide clear and directional instructions, the likelihood of correct image cropping significantly improves. Specifically, GPT-4o achieves 100\% Crop Recall when considering only requests with explicitly described target areas.
Moreover, GPT-4o exhibits shorter generation times compared to smaller edge-based models. This efficiency stems from the superior processing capabilities of cloud servers and the fact that smaller models struggle with structured output generation, often requiring multiple iterations that increase latency.

Evaluating the refined stage requires subjective human assessments for some requests (e.g., determining if responses align with user intent). Before a comprehensive subjective evaluation in \Cref{sec:user:study}, we focus on deterministic tasks here, such as generating objects at specific locations and creating animations with targets on specific real-world positions.
An object creation response is deemed successful if it correctly specifies the prefab name, coordinate space (pixel space), and pixel position (within 1.5 times the bounding box of the detected object). Similarly, animation responses must have the correct action type, object name, coordinate space, and pixel position.
Note that although our framework supports three different coordinate spaces, we focus our evaluation solely on requests requiring pixel space, as this best aligns with real-world interactions. Additionally, we apply a 1.5 multiplier to tolerate coarse position descriptions, which accommodates cases where the object is `nearby,' ensuring that both left and right positions are considered valid.
Similar to the initial stage, we propose 10 typical user requests per category (object/animation)
and repeat experiments 10 times to compute averages. As demonstrated in \Cref{tab:eva:accuracy}, results indicate that cloud models continue to outperform edge-based models, with accuracy in the refined stage being higher due to reduced ambiguity in requests. Given its strong performance, GPT-4o is selected for our user studies.

\subsection{Colocation Accuracy and Efficiency}
\sysname employs a semi-marker-based approach to align the pose of virtual objects among users, where alignment accuracy can be evaluated using the spatial inconsistency metric proposed by \cite{ran2020multi}. This metric quantifies the deviation in rendering the same virtual content across multiple users. We register client devices from various locations and angles and compute the positional difference between transformation matrices estimated by XR devices' internal localization system and those derived from AprilTag.

\begin{figure}[hbtp]
    \centering
    \subfloat[Accuracy.]{\includegraphics[height=0.16\textwidth]{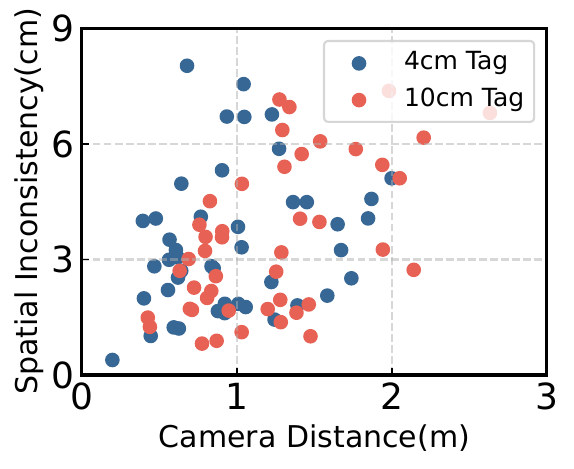}\label{fig:eva:regi:dist}}
    \hfill
    \subfloat[Latency.]{\includegraphics[height=0.16\textwidth]
    {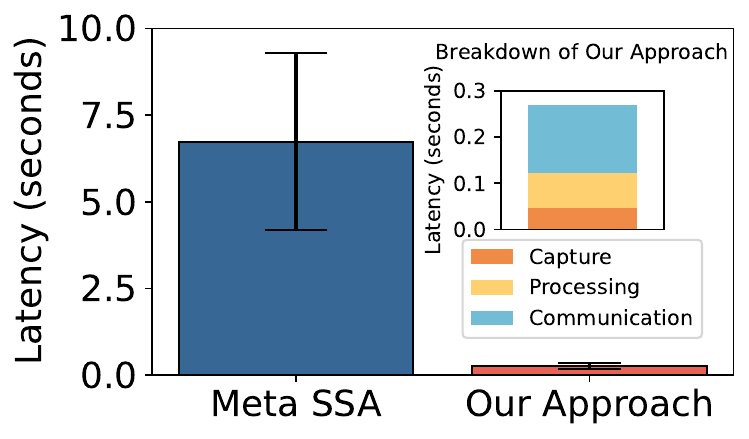}\label{fig:eva:regi:latency}}
    \caption{Registration evaluation.}
    \label{fig:eva:regi}
\vspace{-0.1in}
\end{figure}

To analyze the relationship between camera distance (i.e., the distance between the position where the user registers the device) and spatial inconsistency, we collected data from approximately 70 user pairs using two different tag sizes. The results are demonstrated in \Cref{fig:eva:regi:dist}, where each data point represents a registered user pair.
Our analysis shows that when users register within a 3-meter range, spatial inconsistency remains low, averaging 3.46 cm. In general, closer proximity between users leads to reduced inconsistency. Furthermore, for larger distances, using a larger tag improves performance, as the limited resolution of the captured frames makes it difficult to detect smaller tags at greater distances.
Importantly, once registration is complete, spatial inconsistency does not increase over time, as subsequent localization relies solely on XR devices' internal localization system. For instance, Meta Quest 3 has been reported with a stable tracking of less than 1 cm relative pose error and less than 8 cm absolute pose error through comprehensive measurements \cite{hu2024apple}.

Besides spatial inconsistency, we also evaluate the latency involved in the user registration process, measured from the detection of the user's registration command to the rendering of a wireframe cube. We use Meta SSA as the baseline, which is a popular colocation solution supported by Meta Quest devices. We run a sample colocation app for 10 times and record the initialization time for two users. Our results, as shown in \Cref{fig:eva:regi:latency}, indicate that the total latency for our registration is only around 0.27 seconds on average, almost 25 times faster than 6.73 seconds measured by Meta SSA.

\subsection{Latency Analysis}
\label{sec:latency}
This section focuses on evaluating the end-to-end latency from a user's voice request to system execution and the synchronization of virtual content among multiple users. Ideally, we expect the system to render outcomes within seconds while ensuring that all users in the same physical space perceive virtual object creation and modification simultaneously. However, due to network and processing latencies, there is a discrepancy between the user initiating an action and others observing the resulting changes. To comprehensively analyze system latency, we measure the following key metrics:
\begin{itemize}
    \item \textbf{Speak-to-action latency}: The time from the end of a user's voice recording to when the refined response is executed, such as rendering virtual objects or animations.
    \item \textbf{Response synchronization latency}: The time elapsed between when MLLM responses are executed by the owner and when the synchronized response is rendered for other users.
    \item \textbf{Interaction synchronization latency}: The time difference between when one user interacts with a virtual object and when other users observe the changes applied to that object.
\end{itemize}

Since clock synchronization across different devices can result in discrepancies exceeding 100 ms \cite{mills2006computer}, we introduce a pseudo user to accurately measure synchronization latency. This pseudo user is co-located with a real user on the same device, receives the forwarded messages from the server, and shares an identical timeline with the real user. Then, a real user operated \sysname in a real-world environment for approximately 20 minutes, freely issuing voice requests related to their surroundings while system latencies were recorded. 
During the experiment, the system achieves an average frame rate of $72.34$ frames per second (FPS). Additionally, we collected around 40 samples for speak-to-action and response synchronization latencies and approximately 40,000 samples for interaction synchronization latency. The disparity in sample size arises because the first two latency types depend on user-generated requests, typically a few per minute, while interaction synchronization latency is captured at a frequency of 60Hz per object. We showcase the mean and standard deviation results for each type of latency in \Cref{tab:sync_lat}. 
It is important to note that speak-to-action latency excludes the time taken for screenshot capture and image processing via the object detection model on the edge server. Since these processes are initiated when the user starts speaking, they typically complete before the voice recording ends. Additionally, local processing time includes contextual data collection, raycasting, and any file I/O or serialization on the edge server and client. The total system processing time is computed without accounting for user confirmation, as this varies from person to person.

\begin{table}[hbtp]
    \centering
    \vspace{-0.1in}
    \begin{tabular}{@{}m{2cm}|c|c@{}}
    \toprule
    \multicolumn{2}{c}{Latency Type} & Value (s) \\
    \midrule
    \multirow{8}{*}{\shortstack[l]{Speak-to-action \\ for Single User \\ Request}}   
       & Transcription & $0.453 \pm 0.188$ \\
       & Initial Stage & $1.905 \pm 1.125$ \\
       & User Confirmation & $1.566 \pm 2.760$ \\
       & Text to Speech & $1.549 \pm 1.099$ \\
       & Refined Stage & $3.162 \pm 1.277$ \\
       & Local Processing & $0.520 \pm 0.623$ \\
       & Communication & $0.177 \pm 0.194$ \\
       & Total w/o Confirm & $7.744 \pm 2.086$ \\
       \hline
       \multicolumn{2}{c|}{Response Synchronization} & $1.169 \pm 1.023$ \\
       \hline
       \multicolumn{2}{c|}{Interaction Synchronization} & $0.016 \pm 0.006$ \\
    \bottomrule
    \end{tabular}
    \caption{Latency decomposition of synchronization.}
    \vspace{-0.15in}
    \label{tab:sync_lat}
\end{table}

On average, \sysname processes each user request in approximately $7.744$ seconds, which is a bit long but still acceptable as evidenced by our user study in \Cref{sec:user:study}.
Furthermore, response synchronization latency averages around $1.169$ seconds, which is not particularly noticeable to users. Once an interactable object is created, interaction synchronization latency is as low as $15.53$ milliseconds, a level of responsiveness unachievable through cloud-based synchronization mechanisms.

\section{Preliminary User Study}\label{sec:user:study}
Besides the numerical study of the system performance, we also want to evaluate the subjective user experience with human participants. We would also like to collect feedback from users to indicate directions for further improvements. In this section, we conduct a preliminary user study and analyze the results for a more comprehensive evaluation of our system's usability and practicality.

\subsection{Procedure}
We have recruited $28$ participants ($13$ females, $15$ males; mean age: $27.57$ years) as $14$ pairs of groups for our user study\footnote{All study procedures were conducted in accordance with the ethical guidelines of the Pennsylvania State University Institutional Review Board (IRB) and were approved under protocol number STUDY00026260.}.
The number of participants is selected to follow guidelines for ``debugging" tests as recommended in prior work (see \cite{bevan2003magic}).
The participants, including students and employees, represent diverse academic majors, including engineering (50\%), science (29\%), and liberal arts (21\%).
This participant profile focuses on individuals innovating in education and gaming with a diverse range of backgrounds, which aligns with our framework's target users.
Among the 28 participants, the majority are novices in XR (64\%), with fewer identifying as intermediate (25\%) or expert (11\%). In terms of software programming experience, participants are more evenly spread: 29\% had less than 1 year, 25\% had over 5 years, and the rest had 1--4 years. For game engine experience, 54\% had less than 1 year, another 32\% had 1--2 years, and only 14\% had more than 3 years. This distribution reflects a diverse representation of participants' backgrounds.

Each user study session takes around one hour and follows the same pattern. At the beginning of the user study, participants are provided with a concise system overview of our motivation and what can be done via using our system. Besides, they are asked to classify a series of objects in the environment into different privacy levels, as detailed in \Cref{sec:privacy-metric}. This is followed by a 10-minute training session, familiarizing them with the basic operations to express requests like creating a cube, coloring it, and moving it. Subsequently, participants are requested to provide demographic information and complete a series of tasks in order. At the end of completing each task, they are asked to fill out the NASA Task Load Index (TLX) \cite{hart1988development} form, which measures the perceived workload.
Once all tasks are finished, the participants have 10 more minutes to freely explore the system by declaring any requests they are interested in. After the experiments, the participants are asked to fill out a questionnaire with a few Likert-scale questions and answer a few open-ended questions with a semi-structured interview. 
The details of the user study design are as follows:
\begin{itemize}
    \item Pre-defined user tasks. We define the following tasks targeting object creation, animation creation, scene understanding, and collaborative drawing. Each task has a clear objective while allowing flexibility in execution. 
    Users are asked to complete the registration process before working on each task, and each group will conduct each task twice with switched roles.
    \begin{enumerate}
        \setlength{\itemsep}{2pt}  
        \setlength{\parskip}{0pt}  
        \item Object creation with real-world input. The first user creates a virtual object (selected from a candidate list, including simple objects like a cube, sphere, cup, keyboard, etc.) attached to a real-world object. The other user changes the color of the object to be similar to an existing object in the real-world environment.
        \item Animation creation with real-world input. The first user creates a virtual object. The other user moves the object to a specific real-world position, e.g., next to a chair.
        \item Fine-grained scene understanding. The first user lets the agent describe the objects in her view.
        The other user lets the agent provide information on real-world objects, e.g., provide a brief introduction of a book. 
        \item Collaborative drawing. The first user creates a whiteboard with a marker and an eraser. Both users work on drawing and erasing content. After drawing, the second user asks the agent to recognize what they have drawn.
    \end{enumerate}
    \item Questionnaire. We propose a questionnaire based on existing surveys \cite{lewis1995ibm,lund2001measuring,chen2025llmer}, which evaluates the system from various aspects, including immersion, usability, interactivity, and privacy. The questionnaire is in the form of a seven-point Likert scale, ranging from 1 (strongly disagree) to 7 (strongly agree). 
    The statements included in the questionnaire are as follows:
    \begin{enumerate}[label={(Q\arabic*)}]
        \setlength{\itemsep}{2pt}  
        \setlength{\parskip}{0pt}  
        \item I felt immersed in using \sysname.
        \item What is produced by the system meets my expectations.
        \item I like the interaction possibilities enabled by the system.
        \item The system was responsive and reacted to my commands in a reasonable time.
        \item It is simple to operate the system.
        \item It is simple to collaborate with the other user.
        \item I felt my privacy was protected when using \sysname.
        \item I would like to use \sysname in the future.
        \item I am pleased with the overall experience.
    \end{enumerate}
    \item Semi-structured interview. We propose several open-ended questions to encourage participants to share their experience after completing all the tasks. The questions cover various aspects of experiences using our system, including usability, privacy protection, collaboration experience, and expectation.
    Besides, the participants are free to share their thoughts not covered in those questions. This step collects realistic feedback from users, either in satisfaction or critiques. While there is some intentional overlap between the questionnaire and the semi-structured interview, the semi-structured interviews are designed to elicit in-depth, qualitative insights that cannot be expressed through scaled responses alone.
    
\end{itemize}

\subsection{Quantitative Results and Discussion}
\subsubsection{Privacy Metric} \label{sec:privacy-metric}
To evaluate the effectiveness of our system in preserving user privacy, we introduced 12 categories of objects into the experimental environment: Cup, Desk, Chair, Book, Bag, Laptop/Phone, Monitor, Laboratory equipment and instruments, Human face, Medicine/Medical record, ID card/Driver's license, and Mail/Personal letter/Statement. Participants were asked to classify these objects into three privacy levels: insensitive, maybe sensitive, and highly sensitive.
This classification allows us to evaluate how effectively our system filters content based on users’ perceived privacy risks.
For each frame captured upon the user's request, we recorded the presence of these objects both before and after applying our privacy-aware frame processing approach.

\begin{table}[hbtp]
\centering
\begin{tabular}{l|c|c}
\hline
Privacy Level & Original Frames & Cropped Frames \\
\hline
Insensitive  & 96.43\% & 50.00\% \\
Maybe sensitive  & 82.14\% & 53.57\% \\
Highly sensitive & 53.57\% & 7.14\% \\
\hline
\end{tabular}
\label{tab:privacy-metric}
\caption{Frames including at least one object at each privacy level.}
\end{table}
\vspace{-0.1in}

As shown in \Cref{tab:privacy-metric}, our approach significantly reduces the presence of highly sensitive objects, lowering the percentage from 53.57\% to 7.14\%, thereby effectively preventing such content from being uploaded to the cloud. Edge cases typically involve overlap between a sensitive object and the user's intended target, e.g., a credit card left on top of a book the user wants to interact with. Recall that in these cases, our system provides a final safeguard by prompting the user to manually confirm whether the cropped frame should be uploaded.
While the percentage of frames containing insensitive and maybe sensitive objects remains relatively high, this is often necessary, as the tasks require interaction with those objects. Nevertheless, the observed reductions still demonstrate that our approach successfully filters out irrelevant content, helping MLLMs focus on the target object and reducing potential privacy leakage.

\subsubsection{Usability Metric}
\begin{figure}[hbtp]
    \centering
    \includegraphics[width=0.95\linewidth]{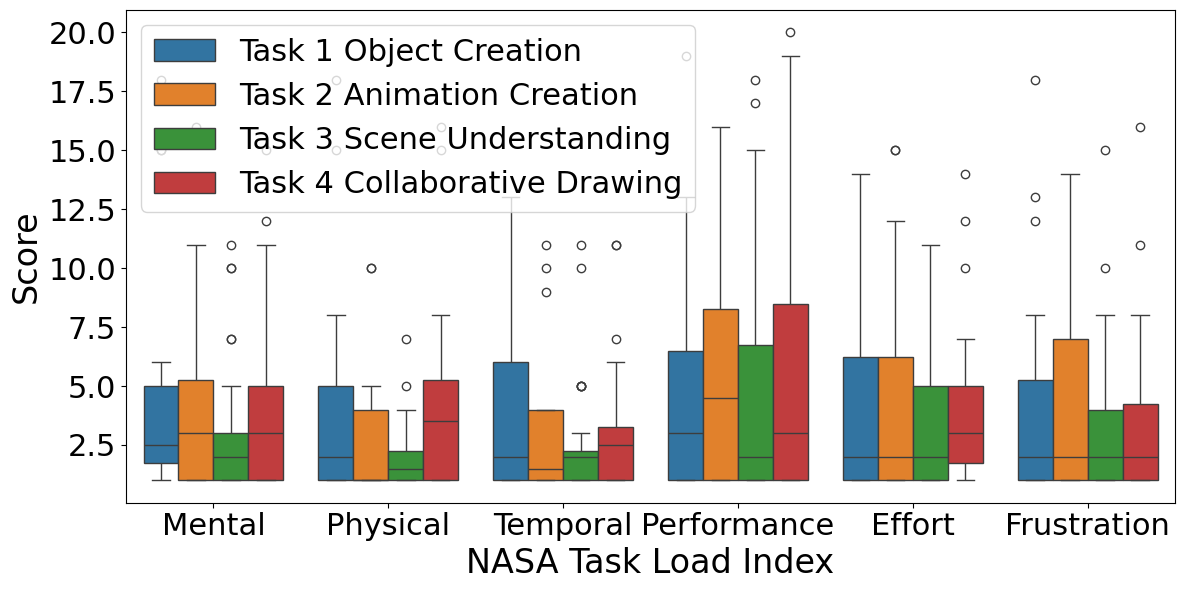}
    \vspace{-0.1in}
    \caption{NASA-TLX scores for all tasks.}
    \vspace{-0.1in}
    \label{fig:nasa-tlx}
\end{figure}

The NASA-TLX scores for each task and metric are presented as a box and whisker plot in \Cref{fig:nasa-tlx}, where the range of the score is from 1 to 20, and lower scores indicate less perceived workload. Across all tasks and metrics, the scores remain relatively low (all average scores are lower than 6/20), suggesting that users can effectively leverage our system for seamless collaboration and interaction with real-world environments.

We also measured the task completion time and task fulfillment level statistics, as shown in \Cref{tab:tct} and \Cref{fig:fulfill}, respectively. Specifically, task completion time is measured from the time the user begins working on a task to the time the user indicates its completion. Note that we do not limit the number of requests a user can issue for a task. They can either integrate all commands together or split them into smaller requests. The fulfillment for each task has four levels: success, minor problem, major problem, and failure, following the definition in \cite{chen2025llmer}.
The results indicate that users spend similar time ($\sim$2 mins) on completing Tasks 1-3 but a longer time on Task 4. It is due to that users usually spend more time thinking about the random drawing or writing on the whiteboard and altering the content before uploading to the server.
Additionally, over 91\% of fulfillment levels are reported as success or minor issues, implying the ease of system usability regardless of the user's background.

\begin{minipage}[t]{0.22\textwidth}\vspace{-0.1in}
    \begin{table}[H] 
    \centering
    \begin{tabular}{@{}lcc@{}}
    \toprule
    Item & Mean (s) & SD (s) \\
    \midrule
    Task 1 & 129.31 & 67.58 \\
    Task 2 & 127.16 & 90.41 \\
    Task 3 & 137.41 & 73.18 \\
    Task 4 & 203.96 & 117.68 \\
    \bottomrule
    \end{tabular}
    \vspace{0.1in}
    \caption{TCT statistics.}
    \label{tab:tct}
    \end{table}
\end{minipage}%
\hfill
\begin{minipage}[t]{0.235\textwidth}\vspace{-0.1in}
    \begin{figure}[H]
    \centering
    \includegraphics[height=0.54\linewidth]{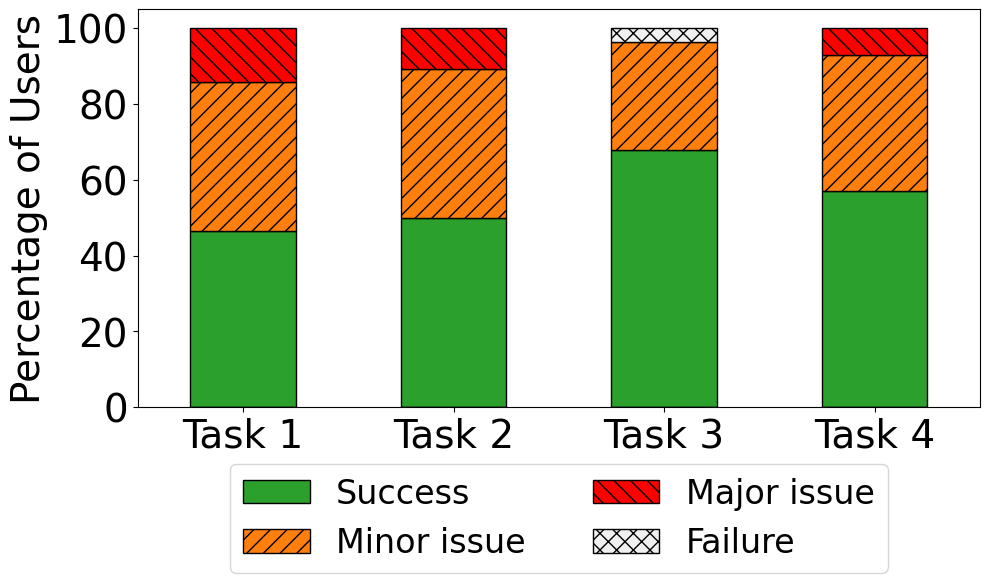}
    \vspace{-0.1in}
    \caption{Fulfillment levels.}
    \label{fig:fulfill}
    \end{figure}
\end{minipage}

\begin{figure}[hbtp]
    \vspace{-0.1in}
    \centering
    \includegraphics[width=0.47\textwidth]{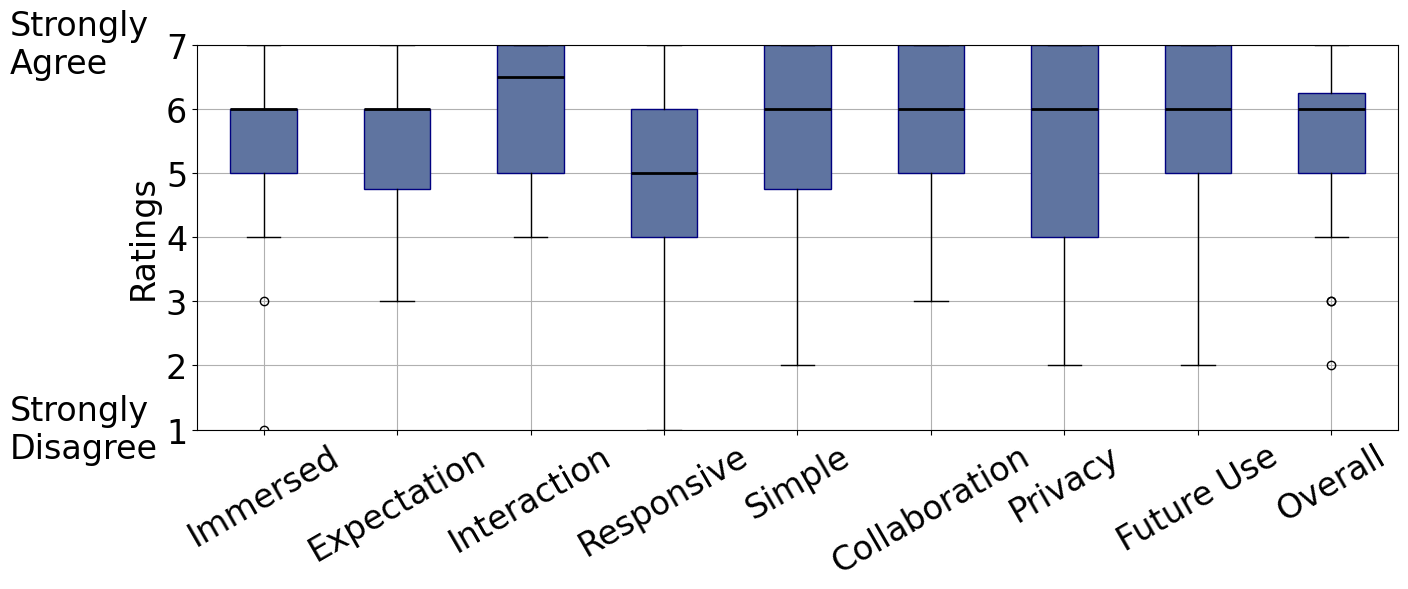}
    \vspace{-0.2in}
    \caption{Results of questionnaire.}
    \vspace{-0.2in}
    \label{fig:questionnaire}
\end{figure}

Similar to NASA-TLX scores, the results of the questionnaire responses are visualized as a box and whisker plot in \Cref{fig:questionnaire}. The system receives generally positive feedback, with median scores ranging from 5 to 7 across all questions.
In particular, user privacy preservation and collaboration were highly rated, receiving average scores of 5.6/7 and 5.8/7, respectively. These results highlight users’ confidence in the system’s privacy-aware frame processing and synchronization mechanisms. On the other hand, the responsiveness yields the lowest average score of 5.0/7, implying the main bottleneck for the overall user experience.

\subsection{User Feedback and Discussion}
We have collected 248 comments across 28 users through our interviews and conducted a thematic analysis \cite{clarke2017thematic} to better understand user experiences with our system.

\subsubsection{Positive Experiences}
\textbf{Immersive collaboration}.
Many participants expressed strong satisfaction with the immersive XR experience and the system's support for interactive, multi-user collaboration. Users highlighted how engaging it was to work with others in a shared virtual space. This theme was frequently mentioned and reflects the system's success in fostering a natural and co-located XR interaction.

\textbf{Perceived privacy protection}.
Another prominent theme was appreciation for the system’s privacy-aware design. Several users positively noted the mechanisms that preserve personal data, and some even proposed additional strategies. One participant suggested, ``You may create virtual objects to occlude sensitive objects in case they cannot be excluded,” indicating user engagement with the idea of extending privacy protections through virtual occlusion.

\textbf{Seamless synchronization}.
Participants also praised the smooth synchronization of virtual content across users. This theme reflects users’ recognition of technical reliability in multi-user content sharing and matches our low interaction synchronization latency in \Cref{tab:sync_lat}. As one user remarked, “It is amazing to see how smoothly the virtual cube reacts to another user’s hand gestures.” 

\subsubsection{Limitations and Areas for Improvement}
\textbf{Further privacy improvement.} While the system currently requires users to manually confirm before uploading cropped images to the cloud, feedback indicates that users may sometimes confirm out of habit without reviewing the content. To address this, participants suggested adding text warnings that list the detected objects or visually boxing them within the cropped image.
Moveover, a supplementary approach in those cases with overlapped sentive and insensitive objects would be to apply dynamic occlusion techniques, such as overlaying a virtual mask or covering over the sensitive object, and uploading the partially occluded frame instead.
These enhancements would encourage more mindful confirmation and further strengthen privacy protection.

\textbf{Voice recognition robustness.}
One frequently reported issue concerned the wake word activation system. Participants noted that the keyword recognition was sensitive to variations in accent or tone and sometimes triggered incorrectly. 
Moreover, the transcription of users' voice into text requests is not perfect, while an inaccurate translation will lead to the system executing in the wrong direction.
This theme indicates a need for more reliable and user-friendly voice interaction mechanisms. With the integration of more advanced acoustic processing techniques, the system could further enhance recognition accuracy and overall user experience.

\textbf{Registration process reliability}.
Several users reported difficulties with the lightweight registration process. Although each attempt required minimal time, some participants found it frustrating to repeat the process multiple times to achieve precise alignment with their collaborators.
This finding underscores the importance of a more robust registration mechanism, particularly under conditions involving occlusion or varying lighting. Future work could explore the integration of advanced visual feature extraction and localization algorithms to improve coordinate space alignment and maintain a seamless user experience.

\textbf{High response latency.}
As discussed in \Cref{sec:latency}, the total speak-to-action latency in \sysname averages around 7.7 seconds. While this delay was deemed acceptable by most users during the study, it remains relatively high for real-time XR applications. The primary bottleneck lies in cloud processing time, which is affected by shared usage load on commercial platforms and the inference latency of large MLLMs. Future efforts could focus on deploying smaller, open-source MLLMs on the edge server to reduce latency while maintaining sufficient capability. This shift could also enhance privacy and reduce operational costs.

\textbf{Limited scalability and scope of user study.}
The current user study involved only dyadic interactions and short, predefined tasks. To better assess real-world applicability and system stability, future research should include longer-term or large-scale evaluations involving more users and complex environments. Additionally, certain user interface elements may benefit from further refinement. For instance, the current confirmation mechanism using voice or gesture inputs may lead to inattentive ``click-through'' behavior. Dedicated studies are needed to identify more reliable alternatives for capturing user confirmations and reactions.

\section{Conclusion}\label{sec:conclusion}
In this paper, we proposed a collaborative and privacy-aware XR platform powered by MLLMs. By integrating edge-assisted frame processing with explicit user consent mechanisms, our system enables fine-grained contextual understanding of real-world environments while safeguarding user privacy. To facilitate seamless interaction among multiple users in the same physical space, we introduced a semi-marker-based registration process utilizing tag detection and a fully customizable content synchronization mechanism. To assess the usability of our system, we conducted both a system-level performance evaluation and an IRB-approved preliminary user study. The results demonstrated the efficiency of our approach while also identifying key areas for further optimization.

\acknowledgments{
GenAI tools were used to polish the writing in this paper and generate avatars on the teaser image. This work has been supported in part by NSF under the grants CNS-2152658 and M3X-2420351, and DARPA grant HR0011-2420366.}

\bibliographystyle{abbrv-doi}

\bibliography{refs}

\end{document}